\pdfoutput=1
\documentclass{vldb}

\usepackage{graphicx}
\usepackage{balance}  
\usepackage{type1cm}     
\usepackage{algorithm}     
\usepackage{algorithmic}     
\usepackage{xspace}     
\usepackage{booktabs}     
\usepackage[bf,tableposition=top]{caption}     
\usepackage{siunitx}          
\usepackage[hyphens]{url}     
\usepackage[show]{chato-notes}
\usepackage{soul}
\DeclareGraphicsExtensions{.pdf,.png,.jpg,.eps}
\graphicspath{{./img/}{./fig/}{./experiments/}}

\usepackage[square,numbers]{natbib}     
\newcommand{\spara}[1]{\smallskip\noindent\textbf{#1}}

\newenvironment {squishlist}
{\begin{list}{$\bullet$}
  { \setlength{\itemsep}{1pt}
     \setlength{\parsep}{1pt}
     \setlength{\topsep}{1pt}
     \setlength{\partopsep}{1pt}
     \setlength{\leftmargin}{1.5em}
     \setlength{\labelwidth}{1em}
     \setlength{\labelsep}{0.5em} } }
{\end{list}}

\newcommand{\samoa}{\textsc{SAMOA}\xspace}
\newcommand{\asamoa}{Apache \samoa}

\newcommand{\moa}{MOA\xspace}

\newcommand{\vht}{VHT\xspace}

\newcommand{\vfdt}{VFDT\xspace}

\newcommand{\wok}{\textbf{wok}\xspace}
\newcommand{\wk}[1]{\textbf{wk(#1)}\xspace}
\newcommand{\wkz}{\wk{z}}

\begin{document}
\title{VHT: Vertical Hoeffding Tree}
\numberofauthors{1}

\author{
\begin{tabular}{cccc}
Nicolas Kourtellis &  Gianmarco De Francisci Morales & Albert Bifet & Arinto Murdopo  \\
\affaddr{Telefonica Research} & \affaddr{Qatar Computing Research Institute} & \affaddr{Telecom ParisTech} & \affaddr{LARC-SMU} \\
\textsf{nicolas.kourtellis@telefonica.com} & \textsf{gdfm@acm.org} & \textsf{albert@albertbifet.com} & \textsf{arintom@smu.edu.sg} \\
\end{tabular}
}

\maketitle

\begin{abstract}
IoT Big Data requires new machine learning methods able to scale to large size of data arriving at high speed. 
Decision trees are popular machine learning models since they are very effective, yet easy to interpret and visualize.
In the literature, we can find distributed algorithms for learning decision trees, and also streaming algorithms, but not algorithms that combine both features.
In this paper we present the Vertical Hoeffding Tree (\vht), the first distributed streaming algorithm for learning decision trees.
It features a novel way of distributing decision trees via vertical parallelism.
The algorithm is implemented on top of \asamoa, a platform for mining distributed data streams, and thus able to run on real-world clusters.
We run several experiments to study the accuracy and throughput performance of our new \vht algorithm, as well as its ability to scale while keeping its superior performance with respect to non-distributed decision trees.
\end{abstract}

\section{Introduction}
\label{sec:introduction}

Nowadays, we generate data from many of our daily activities as we interact with software systems continuously.
The posts in a social network like Twitter or Facebook, the purchases with a credit card, the clicks in a website, or the access to the GPS, can all potentially produce useful information for interested parties.
The recent advancements in mobile devices and wearable technology have further increased the rate and amount of data being generated.
People now generate data anywhere, anytime, by using a multitude of gadgets and technologies.
In the limit, the Internet of Things (IoT) will continuously produce data without any human intervention, thus leading to a dramatic increase volume and velocity of data.
It is estimated that the IoT will consist of almost 50 billion objects by 2020.\footnote{\url{http://www.cisco.com/c/dam/en_us/about/ac79/docs/innov/IoT_IBSG_0411FINAL.pdf}}

There is a common pattern to most modern data sources: data is generated continuously, as a \emph{stream}.
Extracting knowledge from these massive streams of data to create models, and using them, e.g., to choose a suitable business strategy, or to improve healthcare services, can generate substantial competitive advantages.
Many applications need to process incoming data and react on-the-fly by using comprehensible prediction mechanisms.
For example, when a bank monitors the transactions of its clients to detect frauds, it needs to identify and verify a fraud as soon as the transaction is performed, and immediately either block it, or adjust the prediction mechanism.

Streaming data analytic systems need to process and manage data streams in a fast and efficient way, due to the stringent restrictions in terms of time and memory imposed by the streaming setting.
The input to the system is an unbounded stream arriving at high speed.
Therefore, we need to use simple models that scale gracefully with the amount of data.
Additionally, we need to let the model take the right decision online.
But how can we trust that the model is right?
A way to create trust is to enhance understanding of the model and its interpretability, for instance via visualization.
There are several models which satisfy both requirements, however, for reasons we discuss further, in this work we focus on \emph{decision trees}.

A decision tree is a classic decision support tool that uses a tree-like model.
In machine learning, it can be used for both classification and regression~\citep{Breiman1984cart}.
At its core, a decision tree is a model where internal nodes are tests on attributes, branches are possible outcomes of these tests, and leafs are decisions, e.g., a class assignment.

Decision trees, and in general tree-based classifiers, are widely popular, for several reasons.
First, the model is very easy to interpret.
It is easy to understand how the model reaches a classification decision, and the relative importance of features.
Trees are also easy to visualize, and to modify according to domain knowledge.
Second, prediction is very fast.
Once the model is trained, classifying a new instance requires just a logarithmic number of very fast checks (in the size of the model).
For this reason, they are commonly used in one of the most time-sensitive domains nowadays -- Web search~\citep{Chapelle2011ltrchallenge, tyree2011parallel}.
Third, trees are powerful classifiers that can model non-linear relationships.
Indeed, their performance, especially when used in ensemble methods such as boosting, bagging, and random forests, is outstanding~\citep{Delgado2014comparison}.

Learning the optimal decision tree for a given labeled dataset is NP-complete even for very simple settings~\citep{Hyafil1976optimal}.
Practical methods for building tree models usually employ a greedy heuristic that optimizes decisions locally at each node~\citep{Breiman1984cart}.
In a nutshell, the greedy heuristic starts with an empty node (the root) as the initial model, and works by recursively sorting the whole dataset through the current model.
Each leaf of the tree collects statistics on the distribution of attribute-class co-occurrences in the part of dataset that reaches the leaf.
When all the dataset has been analyzed, each leaf picks the best attribute according to a \emph{splitting criterion} (e.g., entropy or information gain).
Then, it becomes an internal node that branches on that attribute, splits the dataset into newly created children leaves, and calls the procedure recursively for these leaves.
The procedure usually stops when the leaf is pure (i.e., only one class reaches the leaf), or when the number of instances reaching the leaf is small enough.
This recursive greedy heuristic is inherently a batch process, as it needs to process the whole dataset before taking a split decision.
However, streaming variants of tree learners also exist.

The \emph{Hoeffding tree}~\citep{Domingos2000vfdt} (a.k.a. \vfdt) is a streaming decision tree learner with statistical guarantees.
In particular, by leveraging the Chernoff-Hoeffding bound~\citep{Hoeffding1963bound}, it guarantees that the learned model is asymptotically close to the model learned by the batch greedy heuristic, under mild assumptions.

The learning algorithm is very simple.
Each leaf keeps track of the statistics for the portion of the stream it is reached by, and computes the best two attributes according to the splitting criterion.
Let $\Delta G$ be the difference between the value of the functions that represent the splitting criterion of these two attributes.
Let $\epsilon$ be a quantity that depends on a user-defined confidence parameter $\delta$, and that decreases with the number of instances processed.
When $\Delta G > \epsilon$, then the currently best attribute is selected to split the leaf.
The Hoeffding bound guarantees that this choice is the correct one with probability larger than $1 - \delta$.

Streaming algorithms are only one of the two main ways to deal with massive datasets, the other being distributed algorithms~\citep{deFrancisciMorales2013samoa}.
However, even though streaming algorithms are very efficient, they are still bounded by the limits of a single machine.
As argued by~\citet{Agarwal2011vw}, ``there are natural reasons for studying distributed machine learning on a cluster.''
Nowadays, the data itself is usually already distributed, and the cost of moving it to a single machine is too high.
Furthermore, cluster computing with commodity servers is economically more viable than using powerful single machines, as testified by innumerable web companies~\citep{Barroso2009datacenter}.
Finally, ``the largest problem solvable by a single machine will always be constrained by the rate at which the hardware improves, which has been steadily dwarfed by the rate at which our data sizes have been increasing over the past decade''~\citep{Agarwal2011vw}.

For all the aforementioned reasons, the goal of the current work is to propose a tree learning algorithm for the streaming setting that runs in a distributed environment.
By combining the efficiency of streaming algorithms with the scalability of distributed processing we aim at providing a practical tool to tackle the complexities of ``big data'', namely its velocity and volume.
Specifically, we develop our algorithm in the context of \asamoa~\cite{DeFrancisciMorales2015samoa}, an open-source platform for mining big data streams.\footnote{\url{http://samoa.incubator.apache.org}}

We name our algorithm the Vertical Hoeffding Tree (\vht).
The \emph{vertical} part stands for the type of parallelism we employ, namely, vertical data parallelism.
Similarly to the original Hoeffding tree, the \vht features anytime prediction and continuous learning.

Naturally, the combination of streaming and distributed algorithms presents its own unique challenges.
Other approaches have been proposed for parallel algorithms, which however do not take into account the characteristics of modern, shared-nothing cluster computing environments~\citep{ben-haim2010spdt}.

Concisely, we make the following contributions:
\begin{squishlist}
\item we propose \vht, the first distributed streaming algorithms for learning decision trees;
\item in doing so, we explore a novel way of parallelizing decision trees via vertical parallelism;
\item we deploy our algorithm on top of SAMOA, and run it on a real-world Storm cluster to test scalability and accuracy;
\item we experiment with large datasets of tens of thousands of attributes and obtain high accuracy (up to $80\%$) and high throughput (offering up to $20x$ speedup over serial streaming solutions).
\end{squishlist}

The outline of the paper is as follows.
We discuss related work in Section~\ref{sec:related}, and some preliminary concepts in Section~\ref{sec:preliminaries}.
We present the new \vht algorithm in Section~\ref{sec:algorithm}, various optimization and implementation details in Section~\ref{sec:implementation}, and an empirical evaluation in Section~\ref{sec:experiments}, with several experimental setups on real and synthetic datasets.
Finally, with Section~\ref{sec:conclusion} we conclude this work.

\section{Related Work}
\label{sec:related}

The literature abounds with streaming and distributed machine learning algorithms, though none of these features both characteristics simultaneously.
Reviewing all these algorithms is out of the scope of this paper, so we focus our attention on decision trees.
We also review the few attempts at creating distributed streaming learning algorithms that have been proposed so far.

\spara{Algorithms.}
One of the pioneer works in decision tree induction for the streaming setting is the Very Fast Decision Tree algorithm (\vfdt)~\cite{Domingos2000vfdt}. This work focuses on alleviating the bottleneck of machine learning application in terms of time and memory, i.e. the conventional algorithm is not able to process it due to limited processing time and memory. Its main contribution is the usage of the Hoeffding Bound to decide the number of data required to achieve certain level of confidence. 
This work has been the basis for a large number of improvements, such as dealing with concept drift~\citep{GamaZBPB14} and handling continuous numeric attributes~\citep{GamaRM03}. 

PLANET~\cite{panda_planet_2009} is a framework for learning tree models on massive datasets.
This framework utilizes MapReduce to provide scalability.
The authors propose a PLANET scheduler to transform steps in decision tree induction into MapReduce jobs that can be executed in parallel.
PLANET uses task-parallelism where each task (node splitting) is executed by one MapReduce jobs that runs independently.
Clearly, MapReduce is a batch programming paradigm which is not suited to deal with streams of data.

\citet{ye_stochastic_2009} show how to distribute and parallelize Gradient Boosted Decision Trees (GBDT).
The authors first implement MapReduce-based GBDT that employs horizontal data partitioning.
Converting GBDT to MapReduce model is fairly straightforward.
However, due to high overhead from HDFS as communication medium when splitting nodes, the authors conclude that MapReduce is not suitable for this kind of algorithm.
The authors then implement GBDT by using MPI.
This implementation uses vertical data partitioning by splitting the data based on their attributes.
This partitioning technique minimizes inter-machine communication cost.
Vertical parallelism is also the data partitioning strategy we choose for the \vht algorithm.

While technically not a tree, \citet{vu2014distributed} propose the first distributed streaming rule-based regression algorithm.
The algorithm is in spirit similar to the \vht, as it uses vertical parallelism and runs on top of distributed SPEs.
However, it creates a different kind of model and deals with regression rather than classification.

\spara{Frameworks.}
We identify two frameworks that belong to the category of distributed streaming machine learning: Jubatus and StormMOA.
Jubatus\footnote{\url{http://jubat.us/en}} is an example of distributed streaming machine learning framework.
It includes a library for streaming machine learning such as regression, classification, recommendation, anomaly detection and graph mining.
It introduces the local ML model concept which means there can be multiple models running at the same time and they process different sets of data.
Using this technique, Jubatus achieves horizontal scalability via horizontal parallelism in partitioning data.
We test horizontal parallelism in our experiments, by implementing a horizontally scaled version of the hoeffding tree.

Jubatus establishes tight coupling between the machine learning library implementation and the underlying distributed stream processing engine (SPE).
The reason is Jubatus builds and implements its own custom distributed SPE.
In addition, Jubatus does not offer any tree learning algorithm, as all of its models need to be linear by construction.

StormMOA\footnote{\url{http://github.com/vpa1977/stormmoa}} is a project to combine MOA with Storm to satisfy the need of scalable implementation of streaming ML frameworks.
It uses Storm's Trident abstraction and MOA library to implement OzaBag and OzaBoost\cite{oza_online_2001}.

Similarly to Jubatus, StormMOA also establishes tight coupling between MOA (the machine learning library) and Storm (the underlying distributed SPE).
This coupling prevents StormMOA's extension by using other SPEs to execute the machine learning library.

StormMOA only allows to run a single model in each Storm bolt (processor).
This characteristic restricts the kind of models that can be run in parallel to ensembles.
The \emph{sharding} algorithm we use in the experimental section can be seen as an instance of this type of framework.

\section{Preliminaries}
\label{sec:preliminaries}

This section introduces the background needed to understand the \vht algorithm.
First, we review the literature on inducing decision trees on a stream.
Then, we present the programming paradigm offered by \asamoa.

\subsection{Hoeffding Tree}
A decision tree consists of a tree structure, where each internal node corresponds to a test on an attribute.
The node splits into a branch for each attribute value (for discrete attributes), or a set of branches according to ranges of the value (for continuous attributes).
Leaves contain classification predictors, usually majority class classifiers, i.e., each leaf predicts the class belonging to the majority of the instances that reach the leaf.

Decision tree models are very easy to interpret and visualize.
The class predicted by a tree can be explained in terms of a sequence of tests on its attributes.
Each attribute contributes to the final decision, and it's easy to understand the importance of each attribute.\footnote{\url{http://blog.datadive.net/interpreting-random-forests}}

\begin{algorithm}[t]
\caption{HoeffdingTreeInduction($X$, $HT$, $\delta$)}
\begin{algorithmic}[1]
\REQUIRE $X$, a labeled training instance.
\REQUIRE $HT$, the current decision tree.
\STATE Use $HT$ to sort $X$ into a leaf $l$
\STATE Update sufficient statistic in $l$
\STATE Increment $n_l$, the number of instances seen at $l$
\IF{$n_l$ mod $n_{min}$ $= 0$ {\bf and} not all instances seen at $l$ belong to the same class}
\STATE For each attribute, compute $\overline{G}_{l}(X_{i})$
\STATE Let $X_{a}$ be the attribute with highest $\overline{G}_{l}$
\STATE Let $X_{b}$ be the attribute with second highest $\overline{G}_{l}$
\STATE Compute the Hoeffding bound $\epsilon = \sqrt{\frac{R^{2}\ln(1/\delta)}{2n_{l}}}$
\IF{$X_{a} \ne X_{\emptyset}$ {\bf and} ($\overline{G}_{l}(X_{a}) - \overline{G}_{l}(X_{b}) > \epsilon$ {\bf or} $\epsilon < \tau$)}
\STATE Replace $l$ with an internal node branching on $X_{a}$
\FORALL{branches of the split}
\STATE Add a new leaf with derived sufficient statistic from the split node
\ENDFOR
\ENDIF
\ENDIF
\end{algorithmic}
\label{alg:ht}
\end{algorithm}

The Hoeffding tree or \vfdt is a very fast decision tree for streaming data.
Its main characteristic is that rather than reusing instances recursively down the tree, it uses them only once.
Algorithm~\ref{alg:ht} shows a high-level description of the Hoeffding tree.

At the beginning of the learning phase, it creates a tree with only a single node.
The Hoeffding tree induction Algorithm~\ref{alg:ht} is invoked for each training instance $X$ that arrives.
First, the algorithm sorts the instance into a leaf $l$ (line 1).
This leaf is a \emph{learning leaf}, therefore the algorithm updates the sufficient statistic in $l$ (line 2).
In this case, the sufficient statistic is the class distribution for each value of each attribute.
In practice, the algorithm increases a counter $n_{ijk}$, for attribute $i$, value $j$, and class $k$.
The algorithm also increments the number of instances ($n_l$) seen at leaf $l$ based on $X$'s weight (line 3).

A single instance usually does not change the distribution significantly enough, therefore the algorithm tries to grow the tree only after a certain number of instances $n_{min}$ has been sorted to the leaf.
In addition, the algorithm does not grow the tree if all the instances that reached $l$ belong to the same class (line 4).

To grow the tree, the algorithm attempts to find a good attribute to split the leaf on.
The algorithm iterates through each attribute and calculates the corresponding splitting criterion $\overline{G}_{l}(X_{i})$ (line 5).
This criterion is an information-theoretic function, such as entropy or information gain, which is computed by making use of the counters $n_{ijk}$.
The algorithm also computes the criterion for the scenario where no split takes places ($X_{\emptyset}$).
\citet{Domingos2000vfdt} refer to this inclusion of a no-split scenario with the term \emph{pre-pruning}. 

The algorithm then chooses the best ($X_{a}$) and the second best ($X_{b}$) attributes based on the criterion (lines 6 and 7).
By using these chosen attributes, it computes the difference of their splitting criterion values $\Delta \overline{G}_{l} = \overline{G}_{l}(X_{a}) - \overline{G}_{l}(X_{b})$.
To determine whether the leaf needs to be split, it compares the difference $\Delta \overline{G}_{l}$ to the Hoeffding bound $\epsilon$ for the current confidence parameter $\delta$ (where $R$ is the range of possible values of the criterion).
If the difference is larger than the bound ($\Delta \overline{G}_{l} > \epsilon$), then $X_{a}$ is the best attribute with high confidence $1-\delta$, and can therefore be used to split the leaf.

Line 9 shows the complete condition to split the leaf.
If the best attribute is the no-split scenario ($X_{\emptyset}$), the algorithm does not perform any split.
The algorithm also uses a tie-breaking $\tau$ mechanism to handle the case where the difference in splitting criterion between $X_{a}$ and $X_{b}$ is very small.
If the Hoeffding bound becomes smaller than $\tau$ ($\Delta \overline{G}_{l} <  \epsilon < \tau$), then the current best attribute is chosen regardless of the values of $\Delta \overline{G}_{l}$.

If the algorithm splits the node, it replaces the leaf $l$ with an internal node.
It also creates branches based on the best attribute that lead to newly created leaves and initializes these leaves using the class distribution observed at the best attribute the branches are starting from (lines 10 to 13).

\subsection{\samoa}
\asamoa\footnote{\url{https://samoa.incubator.apache.org}} is an open-source distributed stream mining platform initially developed at Yahoo Labs~\citep{deFrancisciMorales2013samoa}.
It allows easy implementation and deployment of distributed streaming machine learning algorithms on supported distributed stream processing engines (DSPEs).
Besides, it provides the ability to integrate new DSPEs into the framework and leverage their scalability for performing big data mining.

\samoa is both a framework and a library.
As a framework, it allows the algorithm developer to abstract from the underlying execution engine, and therefore reuse their code on different engines.
It features a pluggable architecture that allows it to run on several distributed stream processing engines such as Storm,\footnote{\url{https://storm.apache.org}} Samza,\footnote{\url{https://samza.apache.org}} and Flink.\footnote{\url{https://flink.apache.org}}
This capability is achieved by designing a minimal API that captures the essence of modern DSPEs.
This API also allows to easily write new bindings to port \samoa to new execution engines.
\samoa takes care of hiding the differences of the underlying DSPEs in terms of API and deployment.

An algorithm in \samoa is represented by a directed graph of operators that communicate via messages along streams which connect pairs of nodes.
Borrowing the terminology from Storm, this graph is called a \emph{Topology}.
Each node in a Topology is a \emph{Processor} that sends messages through a \emph{Stream}.
A Processor is a container for the code that implements the algorithm.
At runtime, several parallel replicas of a Processor are instantiated by the framework.
Replicas work in parallel, with each receiving and processing a portion of the input stream.
These replicas can be instantiated on the same or different physical computing resources, according to the DSPE used.
A Stream can have a single source but several destinations (akin to a pub-sub system). 
A Topology is built by using a \emph{Topology Builder}, which connects the various pieces of user code to the platform code and performs the necessary bookkeeping in the background.

A processor receives \emph{Content Events} via a Stream.
Algorithm developers instantiate a Stream by associating it with exactly one source Processor.
When the destination Processor wants to connect to a Stream, it needs to specify the \emph{grouping} mechanism which determines how the Stream partitions and routes the transported Content Events.
Currently there are three grouping mechanisms in \samoa: 
\begin{squishlist}
\item \emph{Shuffle grouping}, which routes the Content Events in a round-robin way among the corresponding Processor instances. This grouping ensures that each Processor instance receives the same number of Content Events from the stream.
\item \emph{Key grouping}, which routes the Content Events based on their \emph{key}, i.e., the Content Events with the same key are always routed by the Stream to the same Processor instance.
\item \emph{All grouping}, which replicates the Content Events and broadcasts them to all downstream Processor instances.
\end{squishlist}

\section{Algorithm}
\label{sec:algorithm}

In this section, we explain the details of our proposed algorithm, the \emph{Vertical Hoeffding Tree}, which is a data-parallel, distributed version of the Hoeffding tree described in Section~\ref{sec:preliminaries}.
First, we describe the parallelization and the ideas behind our design choice.
Then, we present the engineering details and optimizations we employed to obtain the best performance.

\subsection{Vertical Parallelism}
\label{sec:vertical_parallelism}

\begin{figure}
	\centering
	\includegraphics[width=\columnwidth]{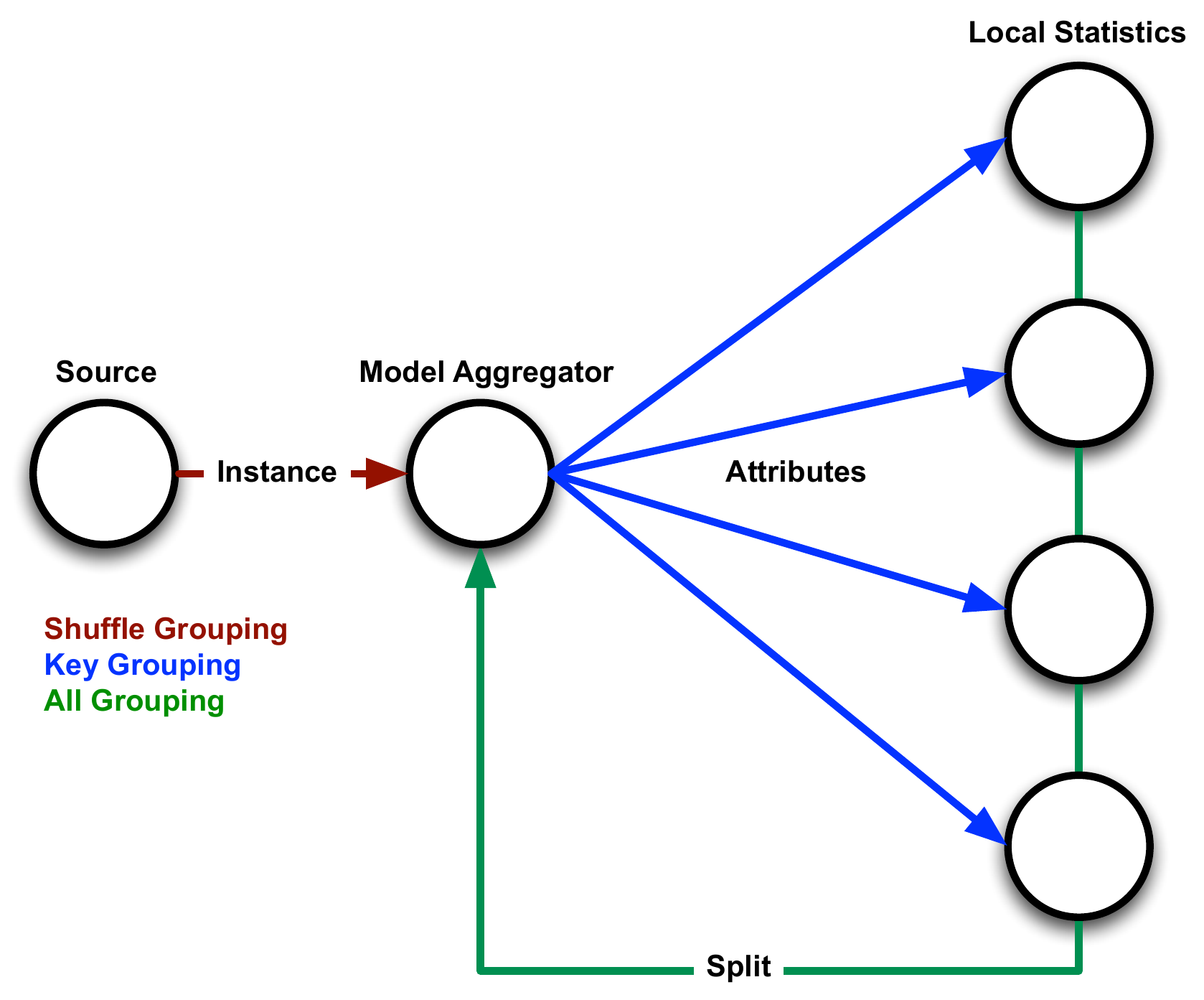}
	\caption{High level diagram of the \vht topology.}
	\label{fig:vht-topology}
\end{figure}

Data parallelism is a way of distributing work across different nodes in a parallel computing environment such as a cluster.
In this setting, each node executes the same operation on different parts of the dataset.
Contrast this definition with task parallelism (aka pipelined parallelism), where each node executes a different operator and the whole dataset flows through each node at different stages.
When applicable, data parallelism is able to scale to much larger deployments, for two reasons:
($i$) data has usually much higher intrinsic parallelism that can be leveraged compared to tasks,
and ($ii$) it is easier to balance the load of a data-parallel application compared to a task-parallel one.
These attributes have led to the high popularity of the currently available DSPEs.
For these reasons, we employ data parallelism in the design of \vht.

In machine learning, it is common to think about data in matrix form.
A typical linear classification formulation requires to find a vector $x$ such that $A \cdot x \approx b$, where $A$ is the data matrix and $b$ is a class label vector.
The matrix $A$ is $n \times m$-dimensional, with $n$ being the number of data instances and $m$ being the number of attributes of the dataset.

Clearly, there are two ways to \emph{slice} this data matrix to obtain data parallelism: by row or by column.
The former is called \emph{horizontal parallelism}, the latter \emph{vertical parallelism}.
With horizontal parallelism, data instances are independent from each other, and can be processed in isolation while considering all available attributes.
With vertical parallelism, instead, attributes are considered independent from each other.

The fundamental operation of the algorithm is to accumulate statistics $n_{ijk}$ (i.e., counters) for triplets of attribute $i$, value $j$, and class $k$, for each leaf of the tree.
The counters for each leaf are independent, so let us consider the case for a single leaf.
These counters, together with the learned tree structure, form the state of the \vht algorithm.

Different kinds of parallelism distribute the counters across computing nodes in different ways.
When using horizontal parallelism, the instances are distributed randomly, therefore multiple instances of the same counter need to exist on several nodes.
On the other hand, when using vertical parallelism, the counters for one attribute are grouped on a single node.

This latter design has several advantages.
First, by having a single copy of the counter, the memory requirements for the model are the same as in the sequential version.
In contrast, with horizontal parallelism a single attribute may be kept track on every node, thus the memory requirements grow linearly with the parallelism level.
Second, by having each attribute being tracked independently, the computation of the split criterion can be performed in parallel by several nodes.
Conversely, with horizontal partitioning the algorithm needs to (centrally) aggregate the partial counters before being able to compute the splitting criterion.

Of course, the vertically-parallel design has also its drawbacks.
In particular, horizontal parallelism achieves a good load balance much more easily, even though solutions for these problems have recently been proposed~\citep{Uddin2015pobc,Uddin2016twochoices}.
In addition, if the instance stream arrives in row-format, it needs to be transformed in column-format, and this transformation generates additional CPU overhead at the source.
Indeed, each attribute that constitutes an instance needs to be sent independently, and needs to carry the class label of its instance.
Therefore, both the number of messages and the size of the data transferred increase.

Nevertheless, as shown in Section~\ref{sec:experiments}, the advantages of vertical parallelism outweigh its disadvantages for several real-world settings.


\subsection{Algorithm Structure}
We are now ready to explain the structure of the \vht algorithm.
Recall from Section~\ref{sec:preliminaries} that there are two main parts to the Hoeffding tree algorithm: \emph{sorting} the instances through the current model, and accumulating \emph{statistics} of the stream at each leaf node.
This separation offers a neat cut point to modularize the algorithm in two separate components.
We call the first component \emph{model aggregator}, and the second component \emph{local statistics}.
Figure~\ref{fig:vht-topology} presents a visual depiction of the algorithm, specifically, of its components and of how the data flow among them.

The model aggregator holds the current model (the tree) produced so far.
Its main duty is to receive the incoming instances and sort them to the correct leaf.
If the instance is unlabeled, the model predicts the label at the leaf and sends it downstream (e.g., for evaluation).
Otherwise, if the instance is labeled it is used as training data.
The \vht decomposes the instance into its constituent attributes, attaches the class label to each, and sends them independently to the following stage, the \emph{local statistics}.
Algorithm~\ref{alg:vht_tree_induction} is a pseudocode for the model aggregator.

\begin{algorithm}[t]
\caption{Model Aggregator: Vertical\-Hoeffding\-TreeInduction($E$, $VHT\_tree$)}
\begin{algorithmic}[1]
\REQUIRE $E$ is a training instance from source PI, wrapped in \verb;instance; content event
\REQUIRE $VHT\_tree$ is the current state of the decision tree in model-aggregator PI
\STATE Use $VHT\_tree$ to sort $E$ into a leaf $l$
\STATE Send \verb;attribute; content events to local-statistic PIs
\STATE Increment the number of instances seen at $l$ (which is $n_l$)
\IF{$n_l$ $mod$ $n_{min}$ $= 0$ {\bf and} not all instances seen at $l$ belong to the same class}
\STATE Add $l$ into the list of splitting leaves
\STATE Send \verb;compute; content event with the id of leaf $l$ to all local-statistic PIs
\ENDIF
\end{algorithmic}
\label{alg:vht_tree_induction}
\end{algorithm}

The local statistics contain the sufficient statistics $n_{ijk}$ for a set of attribute-value-class triples.
Conceptually, the local statistics Processor can be viewed as a large distributed table, indexed by leaf id (row), and attribute id (column).
The value of the cell represents a set of counters, one for each pair of attribute value and class.
The local statistics simply accumulate statistics on the data sent to it by the model aggregator.
Pseudocode for the update function is shown in Algorithm~\ref{alg:update_local_statistic}.

\begin{algorithm}[t]
\caption{Local Statistic: Update\-Local\-Statistic($attribute$, $local\_statistic$)}
\begin{algorithmic}[1]
\REQUIRE $attribute$ is an \verb;attribute; content event
\REQUIRE $local\_statistic$ is the local statistic, could be implemented as $Table<leaf\_id, attribute\_id>$
\STATE Update $local\_statistic$ with data in $attribute$: attribute value, class value and instance weights
\end{algorithmic}
\label{alg:update_local_statistic}
\end{algorithm}

In SAMOA, we implement vertical parallelism by connecting the model to the statistics via key grouping.
We use a composite key made by the leaf id and the attribute id.
Horizontal parallelism can similarly be implemented via shuffle grouping on the instances themselves.


\spara{Messages.}
During the execution of the \vht, the type of events that are being sent and received from the different parts of the algorithm are summarized in Table~\ref{tab:content-events}.

\begin{table*}[htbp]
\caption{Different type of content events used during the execution of the \vht algorithm.}
\vspace{-\baselineskip}
\begin{center}
\small
\tabcolsep=0.5em
\begin{tabular}{l l l l}
\toprule
Name			&	Parameters								&	From					&	To									\\
\midrule
\verb;instance;		&	$<$ attribute 1, \dots, attribute m, class $C >$		&	Source PI				&	Model-Aggregator PI							\\
\verb;attribute; 		&	$<$ attribute $id$, attribute value, class $C >$		&	Model Aggregator PI		&	Local Statistic PI $id = <$ leaf $id$ + attribute $id >$ \\
\verb;compute;		&	$<$ leaf $id >$								&	Model Aggregator PI		&	All Local Statistic PIs							\\
\verb;local-result;	&	$< X^{local}_{a}, X^{local}_{b} >$				&	Local Statistic PI $id$	&	Model Aggregator PI							\\
\verb;drop;		&	$<$ leaf $id >$								&	Model Aggregator PI		&	All Local Statistic PIs							\\
\bottomrule
\end{tabular}
\end{center}
\label{tab:content-events}
\end{table*}%

\spara{Leaf splitting.}
Periodically, the model aggregator will try to see if the model needs to evolve by splitting a leaf.
When a sufficient number of instances have been sorted through a leaf, it will send a broadcast message to the statistics, asking to compute the split criterion for the given leaf id.
The statistics will get the table corresponding to the leaf, and for each attribute compute the splitting criterion in parallel (e.g., information gain or entropy).
Each local statistic Processor will then send back to the model the top two attributes according to the chosen criterion, together with their scores.
The model aggregator simply needs to compute the overall top two attributes, apply the Hoeffding bound, and see whether the leaf needs to be split.
Refer to Algorithm~\ref{alg:receive_compute_message} for a pseudocode.

\begin{algorithm}[t]
\caption{Local Statistic: Receive\-Compute\-Message($compute$, $local\_statistic$)}
\begin{algorithmic}[1]
\REQUIRE $compute$ is an \verb;compute; content event
\REQUIRE $local\_statistic$ is the local statistic, could be implemented as $Table<leaf\_id, attribute\_id>$
\STATE Get leaf $l$ ID from \verb;compute; content event
\STATE For each attribute $i$ that belongs to leaf $l$ in local statistic, compute $\overline{G}_{l}(X_{i})$
\STATE Find $X^{local}_{a}$, which is the attribute with highest $\overline{G}_{l}$ based on the local statistic
\STATE Find $X^{local}_{b}$, which is the attribute with second highest $\overline{G}_{l}$ based on the local statistic
\STATE Send $X^{local}_{a}$ and $X^{local}_{b}$ using \verb;local-result; content event to model-aggregator PI via \verb;computation-result; stream
\end{algorithmic}
\label{alg:receive_compute_message}
\end{algorithm}

Two cases can arise: the leaf needs splitting, or it doesn't.
In the latter case, the algorithm simply continues without taking any action.
In the former case instead, the model modifies the tree by splitting the leaf on the selected attribute, and generating one new leaf for each possible value of the branch.
Then, it broadcasts a \texttt{drop} message containing the former leaf id to the local statistics.
This message is needed to release the resources held by the leaf and make space for the newly created leaves.
Subsequently, the tree can resume sorting instances to the new leaves.
The local statistics will create a new table for the new leaves lazily, whenever they first receive a previously unseen leaf id.
Algorithm~\ref{alg:receive_local_result} shows this part of the process.
In its simplest version, while the tree adjustment is performed, the algorithm drops the new incoming instances.
We show in the next section an optimized version that buffers them to improve accuracy.

\begin{algorithm}[t]
\caption{Model Aggregator: Receive($local\_result$, $VHT\_tree$)}
\begin{algorithmic}[1]
\REQUIRE $local\_result$ is an \verb;local-result; content event
\REQUIRE $VHT\_tree$ is the current state of the decision tree in model-aggregator PI
\STATE Get correct leaf $l$ from the list of splitting leaves
\STATE Update $X_{a}$ and $X_{b}$ in the splitting leaf $l$ with $X^{local}_{a}$ and $X^{local}_{b}$ from $local\_result$
\IF{$local\_results$ from all local-statistic PIs received or time out reached}
\STATE Compute Hoeffding bound $\epsilon = \sqrt{\frac{R^{2}\ln(1/\delta)}{2n_{l}}}$
\IF{$X_{a} \ne X_{\emptyset}$ {\bf and} ($\overline{G}_{l}(X_{a}) - \overline{G}_{l}(X_{b}) > \epsilon$ {\bf or} $\epsilon < \tau$)}
\STATE Replace $l$ with a split-node on $X_{a}$
\FORALL{branches of the split}
\STATE Add a new leaf with derived sufficient statistic from the split node
\ENDFOR
\STATE Send \verb;drop; content event with id of leaf $l$ to all local-statistic PIs
\ENDIF
\ENDIF
\end{algorithmic}
\label{alg:receive_local_result}
\end{algorithm}

\begin{figure}
	\centering
	\includegraphics[width=\columnwidth]{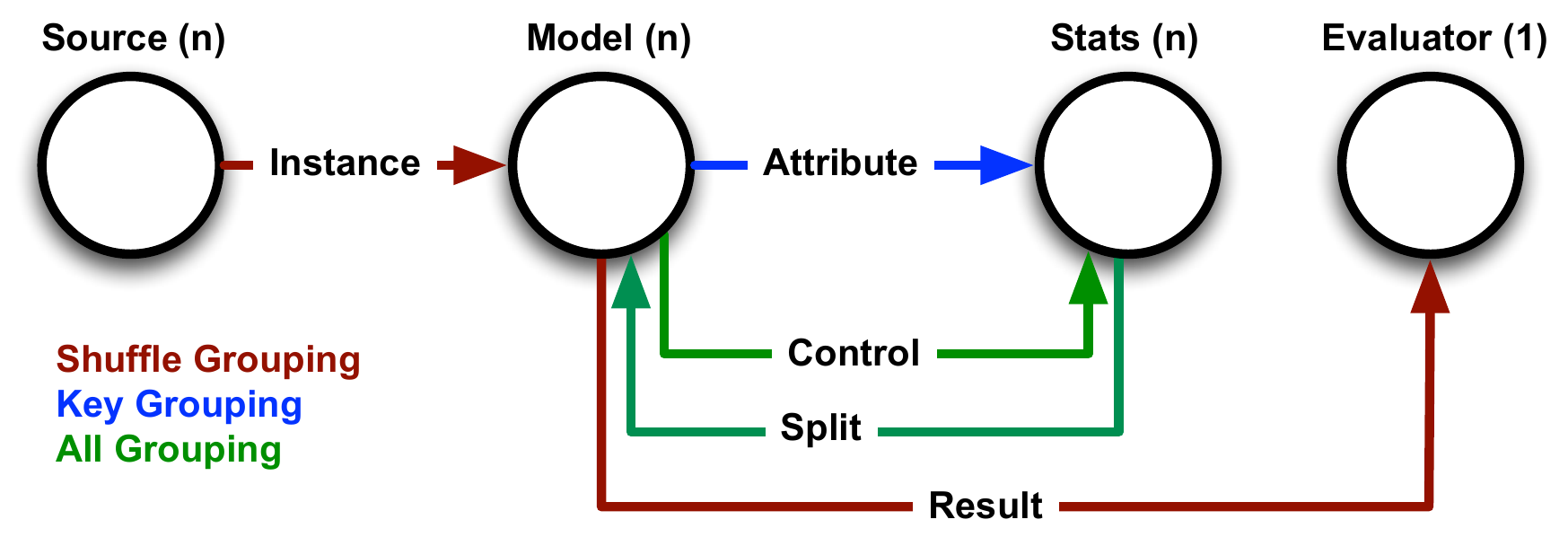}
	\caption{Deployment diagram for \vht.}
	\label{fig:deployment}
\end{figure}

\section{\vht Optimizations}
\label{sec:implementation}

%
%

We now introduce three optimizations that improve the performance of the \vht: \emph{model replication}, \emph{optimistic split execution}, and \emph{instance buffering}.
The first deals with the throughput and I/O capability of the algorithm, by removing its single bottleneck at the model aggregator.
The latter two instead deal with the problem computing the split criterion in a distributed environment.

\spara{Model replication.}
If the model is maintained in a single Processor, it can easily become a bottleneck in the construction and maintenance of the tree, especially under high instance arrival rate.
Instead, a parallel replication and maintenance of the model on multiple Processors allows for higher throughput, but indeed comes with increased management complexity and possible accuracy drop.

In order to materialize the model replication of the \vht, two issues must be resolved: how to distribute incoming instances to models, and how to perform consistent leaf splitting across all models, thus guaranteeing consistent maintenance of the tree in all models.
The first issue can be easily solved via shuffle grouping from the source Processor to all parallel model aggregator Processors.
Assuming $p$ parallel models, shuffle grouping routes incoming instances in a round-robin fashion among them, guaranteeing equal split of instances among the models.

The second issue, however, requires a more elaborate solution, because of two reasons.
First, in a fully distributed mode, each model can decide to send a \emph{control} message to the statistics at any time.
To escape the problem of having inconsistent models, one model (e.g., the first to be created in the topology construction) is appointed the role of the \emph{primary} model and is responsible for broadcasting the \emph{control} message to the statistics.
The frequency of this broadcast is adjusted to take into account the level of model parallelism $p$, and that each model receives $1/p$-th of the total instances.

Second, the exact number of instances $n_l$ seen at each leaf $l$ is not available at any central point.
Instead, each model handles a portion of the stream and thus only a partial number of instances for a leaf $l$ ($n_l'$) is available to each model for the computation of the Hoeffding bound $\epsilon$.
To remedy this problem, a naive approach would be to estimate, at the models, that $n_l \approx n_l' \times p$.
However, this approach can over- or under-estimate the true number of instances seen per leaf, if the instances are not dense, i.e., they don't contain values for all attributes.

A better approach is to have the local statistics broadcast back to all models their estimation of $n_l'$, along with their top two attributes $X^{local}_{a}$ and $X^{local}_{b}$.
Note that, for sparse instances, the value of $n_l'$ is still an estimate and not an exact value due to the way attributes are distributed to statistics via key grouping on the composite key $(leaf\_id,attribute\_id)$.
Indeed, for dense instances, each instance gets decomposed into all its attributes, so the model sends a message per instance to the statistics.
However, for sparse instances, each instance gets decomposed into a subset of its attributes (those with non-zero values).
Therefore, each local statistic may receive a different number of instances per leaf, and thus its counter for $n_l'$ underestimates the real value $n_l$.
However, the models can now independently compute $n_l'' = \max n_l'$, the maximum over all received estimates $n_l'$.
This value is a good estimate of the true value of $n_l$, especially for real-world skewed datasets, where one of the attributes is extremely frequent.
Finally, the models can compute the Hoeffding bound $\epsilon$ for the particular leaf, and decide in a consistent manner across all models whether to split, thus allowing the maintenance of the same tree in all model processors.


\spara{Optimistic split execution.}
In the simplest version of the \vht algorithm, whenever the decision on splitting is being taken, labeled instances are simply thrown away.
This behavior clearly wastes useful data, and is thus not desirable.

Note that there are two possible outcomes when a split decision is taken.
If the algorithm decides to split the current leaf, all the statistics accumulated so far for the leaf are dropped.
Otherwise, the leaf keeps accumulating statistics.
In either case, the algorithm is better served by \emph{using} the instances that arrive during the split.
If the split is taken, the in-transit instances do not have any effect in any case.
However, if the split is not taken, the instances can be correctly used to accumulate statistics.

Given these observations, we modify the \vht algorithm to keep sending instances that arrive during splits to the local statistics.
We call this variant of the algorithm {\wk{0}.

\spara{Instance buffering.}
The feedback for a split  from the local statistics to the model aggregator comes with a delay that can affect the performance of the model.
While the model is waiting to receive this feedback from the local statistics to decide whether a split should be taken, the information from the instances that arrive can be lost if the node splits.
To avoid this waste, we add a buffer to store instances in the model during a split decision.
The algorithm can replay these instances if the model decides to split.
That is, instances that arrive during a split decision are sent downstream and are accounted for in the current local statistics.
If a split occurs, these statistics are dropped, and the instances are replayed from the buffer before resuming with normal operations.
Conversely, if no split occurs, the buffer is simply dropped.

To avoid increasing the memory pressure of the algorithm, the buffer resides on disk.
The access to the buffer is sequential both while writing and when reading, so it does not represent a bottleneck for the algorithm.
We also limit the maximum size of the buffer, to avoid delaying newly arriving instances excessively.
The optimal size of the buffer depends on the number of attributes of the instances, the arrival rate, the delay of the feedback from the local statistics, and the specific hardware configuration.
Therefore, we let the user customize its size with a parameter $z$, and we refer to this version of the algorithm as \wkz.



\spara{Timeout.}
Each model waits for a timeout to receive all responses back from the statistics, before computing the new splits.
This timeout is primarily a system check to avoid the model waiting indefinitely.
This timeout parameter may impact the performance of the tree: if it's too large, many instances will not be stored in the available buffer and therefore lost.
Thus, the size of the buffer is closely related to this timeout parameter, allowing it to have enough instances to be replayed if the model has a leaf that decides to split.



\section{Experiments}
\label{sec:experiments}

In our experimental evaluation of the \vht method, we aim to study the following questions:
\begin{squishlist}
\item[\bf Q1:] How does a centralized \vht compare to a centralized hoeffding tree (available in \moa) with respect to accuracy and throughput?
\item[\bf Q2:] How does the vertical parallelism used by \vht compare to horizontal parallelism?
\item[\bf Q3:] What is the effect of number and density of attributes?
\item[\bf Q4:] How does discarding or buffering instances affect the performance of \vht?
\end{squishlist}

\subsection{Experimental setup}

In order to study these questions, we experiment with five datasets (two synthetic generators and three real datasets), five different versions of the hoeffding tree algorithm, and up to four levels of computing parallelism.
We measure classification accuracy during the execution and at the end, and throughput (number of classified instances per second). 
We execute each experimental configuration ten times, and report the average of these measures.

\spara{Synthetic datasets.}
We use synthetic data streams produced by two random generators: one for dense and one for sparse attributes.

\begin{squishlist}
\item \textbf{Dense attributes} are extracted from a random decision tree.
We test different number of attributes, and include both categorical and numerical types.
The label for each configuration is the number of categorical-numerical used (e.g, 100-100 means the configuration has 100 categorical and 100 numerical attributes).
We produce 10 differently seeded streams with 1M instances for each tree, with one of two balanced classes in each instance, and take measurements every 100k instances.

\item \textbf{Sparse attributes} are extracted from a random tweet generator.
We test different dimensionalities for the attribute space: 100, 1k, 10k.
These attributes represent the appearance of words from a predefined bag-of-words.
On average, the generator produces 15 words per tweet (size of a tweet is Gaussian), and uses a Zipf distribution with skew $z=1.5$ to select words from the bag.
We produce 10 differently seeded streams with 1M tweets in each stream.
Each tweet has a binary class chosen uniformly at random, which conditions the Zipf distribution used to generate the words.
\end{squishlist}

\spara{Real datasets.}
We also test \vht on three real data streams to assess its performance on benchmark data.\footnote{\url{http://moa.cms.waikato.ac.nz/datasets/},\\\url{http://osmot.cs.cornell.edu/kddcup/datasets.html}}
\begin{squishlist}
\item ($elec$) Electricity.
This dataset has 45312 instances, 8 numerical attributes and 2 classes.
\item ($phy$) Particle Physics.
This dataset has 50000 instances for training, 78 numerical attributes and 2 classes.
\item ($covtype$) CovertypeNorm.
This dataset has 581012 instances, 54 numerical attributes and 7 classes.
\end{squishlist}

\spara{Algorithms.}
We compare the following versions of the hoeffding tree algorithm.

\begin{squishlist}
\item \textbf{\moa}: This is the standard Hoeffding tree in \moa.
\item \textbf{local}: This algorithm executes \vht in a local, sequential execution engine. All split decisions are made in a sequential manner in the same process, with no communication and feedback delays between statistics and model.
\item \textbf{\wok}: This algorithm discards instances that arrive during a split decision.
This version is the vanilla \vht.
\item \textbf{\wk{z}}: This algorithm sends instances that arrive during a split decision downstream.
In also adds instances to a buffer of size $z$ until full.
If the split decision is taken, it replays the instances in the buffer through the new tree model.
Otherwise, it discards the buffer, as the instances have already been incorporated in the statistics downstream.
\item \textbf{sharding}: Splits the incoming stream horizontally among an ensemble of Hoeffding trees.
The final prediction is computed by majority voting.
This method is an instance of horizontal parallelism applied to Hoeffding trees.
\end{squishlist}

\spara{Experimental configuration.}
All experiments are performed on a Linux server with 24 cores (Intel Xeon X5650), clocked at 2.67GHz, L1d cache: 32kB, L1i cache: 32kB, L2 cache: 256kB, L3 cache: 12288kB, and 65GB of main memory.
On this server, we run a Storm cluster (v0.9.3) and zookeeper (v3.4.6).
We use \samoa v0.4.0 (development version) and \moa v2016.04 available from the respective project websites.

We use several parallelism levels in the range of $p=2,\dots,16$, depending on the experimental configuration.
For dense instances, we stop at $p=8$ due to memory constraints, while for sparse instances we scale up to $p=16$.
We disable model replication (i.e., use a single model aggregator), as in our setup the model is not the bottleneck.

\subsection{Accuracy and time of \vht local vs. \moa}

In this first set of experiments, we test if \vht is performing as well as its counterpart hoeffding tree in \moa.
This is mostly a sanity check to confirm that the algorithm used to build the \vht does not affect the performance of the tree when all instances are processed sequentially by the model.
To verify this fact, we execute \vht local and \moa with both dense and sparse instances.
Figure~\ref{fig:dense_sparse_moa_vht_accuracy} shows that \vht local achieves the same accuracy as \moa, even besting it at times.
However, \vht local always takes longer than \moa to execute, as shown by Figure~\ref{fig:dense_sparse_moa_vht_times}.
Indeed, the local execution engine of \samoa is optimized for simplicity rather than speed.
Therefore, the additional overhead required to interface \vht to DSPEs is not amortized by scaling the algorithm out.
Future optimized versions of \vht and the local execution engine should be able to close this gap.

\begin{figure}
\centering
\includegraphics[scale=0.8]{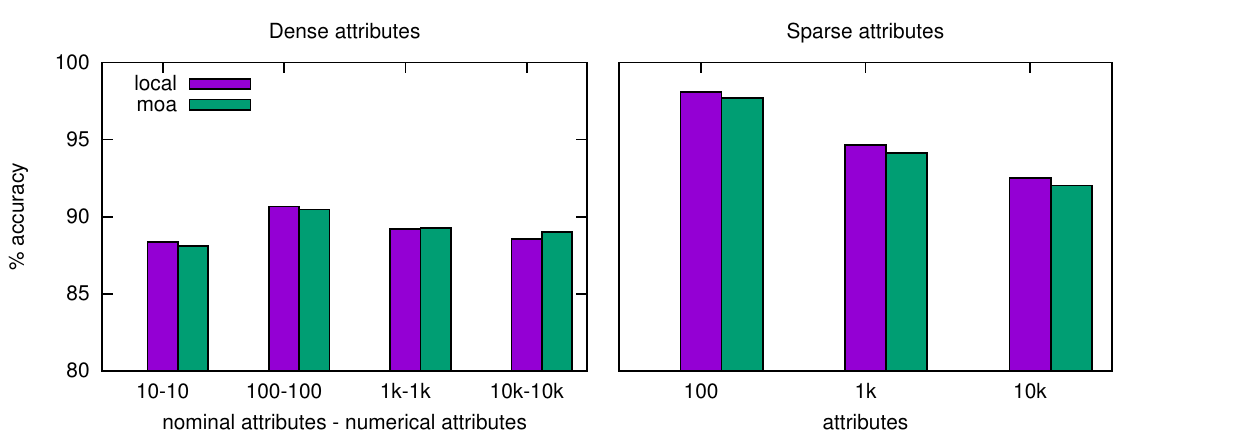}
\caption{Accuracy of \vht executed in local mode on \samoa compared to \moa, for dense and sparse datasets.}
\label{fig:dense_sparse_moa_vht_accuracy}
\end{figure}

\begin{figure}
\centering
\includegraphics[scale=0.8]{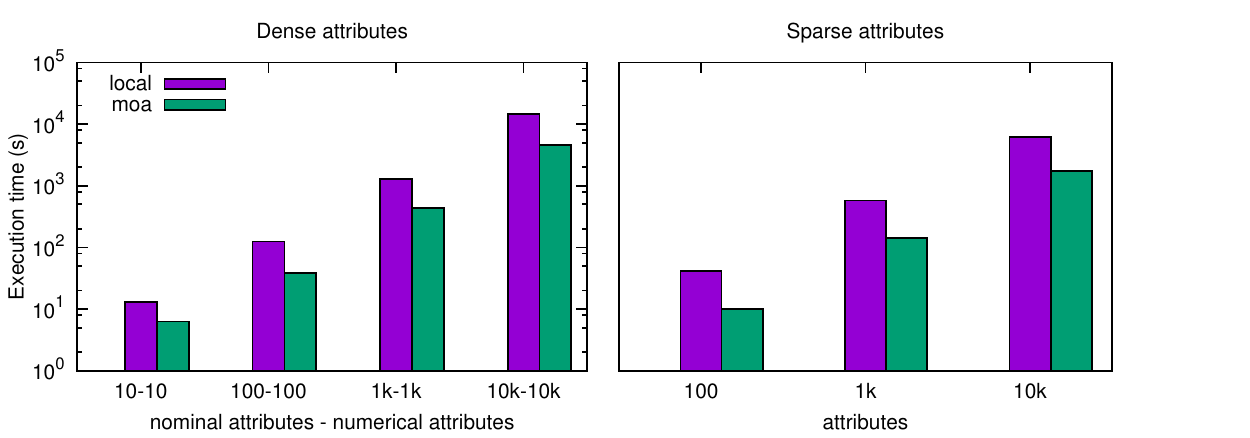}
\caption{Execution time of \vht in local mode on \samoa compared to \moa, for dense and sparse datasets.}
\label{fig:dense_sparse_moa_vht_times}
\end{figure}

\subsection{Accuracy of \vht local vs. distributed}

Next, we compare the performance of \vht local with \vht built in a distributed fashion over multiple processors for scalability.
We use up to $p=8$ parallel statistics, due to memory restrictions, as our setup runs on a single machine.
In this set of experiments we compare the different versions of \vht, \wok and \wkz, to understand what is the impact of keeping instances for training after a model's split.
Accuracy of the model might be affected, compared to the local execution, due to delays in the feedback loop between statistics and model.
That is, instances arriving during a split will be classified using an older version of the model compared to the sequential execution.
As our target is a distributed system where independent processes run without coordination, this delay is a characteristic of the algorithm as much as of the distributed SPE we employ.

We expect that buffering instances and replaying them when a split is decided would improve the accuracy of the model.
In fact, this is the case for dense instances with a small number of attributes (i.e., around $200$), as shown in Figure~\ref{fig:dense_accuracy_withlocal}.
However, when the number of available attributes increases significantly, the load imposed on the model seems to outweigh the benefits of keeping the instances for replaying.
We conjecture that the increased load in computing the splitting criterion in the statistics further delays the feedback to compute the split.
Therefore, a larger number of instances are classified with an older model, thus negatively affecting the accuracy of the tree.
In this case, the additional load imposed by replaying the buffer further delays the split decision.
For this reason, the accuracy for \vht \wkz drops by about $30\%$ compared to \vht local.
Conversely, the accuracy of \vht \wok drops more gracefully, and is always within $18\%$ of the local version.

\vht always performs approximatively $10\%$ better than sharding.
For dense instances with a large number of attributes (20k), sharding fails to complete due to its memory requirements exceeding the available memory.
Indeed, sharding builds a full model for each shard, on a subset of the stream.
Therefore, its memory requirements are $p$ times higher than a standard hoeffding tree.



\begin{figure*}
\centering
\includegraphics[scale=0.8]{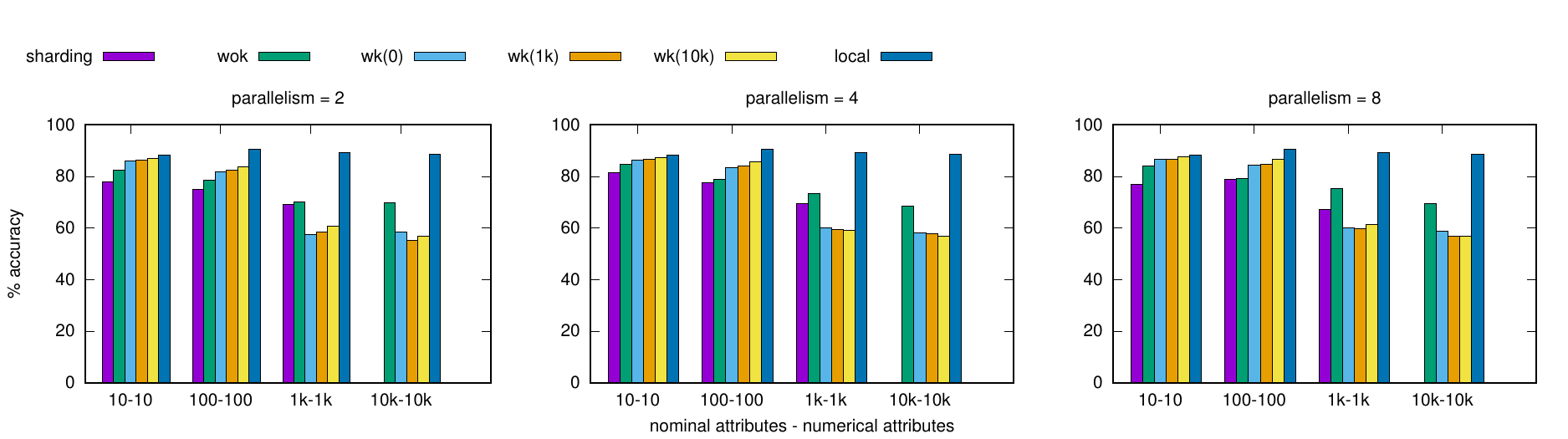}
\caption{Accuracy of several versions of \vht (local, \wok, \wkz) and sharding, for dense datasets.}
\label{fig:dense_accuracy_withlocal}
\end{figure*}

When using sparse instances, the number of attributes per instance is constant, 
while the dimensionality of the attribute space increases.
In this scenario, increasing the number of attributes does not put additional load on the system.
Indeed, Figure~\ref{fig:sparse_accuracy_withlocal} shows that the accuracy of all versions is quite similar, and close to the local one.
This observation is in line with our conjecture that the overload on the system is the cause for the drop in accuracy on dense instances.

\begin{figure*}
\centering
\includegraphics[scale=0.8]{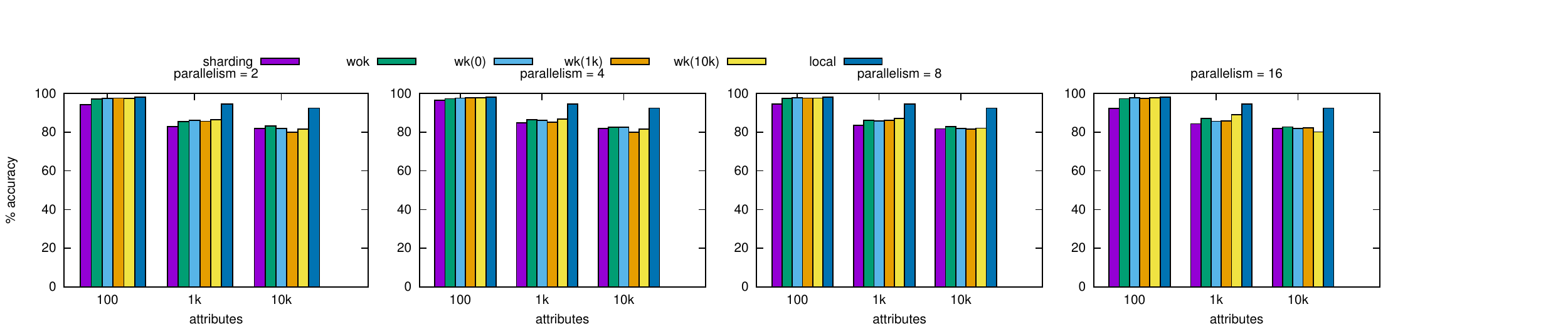}
\caption{Accuracy of several versions of \vht (local, \wok, \wkz) and sharding, for sparse datasets.}
\label{fig:sparse_accuracy_withlocal}
\end{figure*}

We also study how the accuracy evolves over time. 
In general, the accuracy of all algorithms is rather stable, as shown in Figures~\ref{fig:dense_accuracy_evolution} and~\ref{fig:sparse_accuracy_evolution}.
For instances with 10 to 100 attributes, all algorithms perform similarly.
For dense instances, the versions of \vht with buffering, \wkz, outperform \wok, which in turn outperforms sharding.
This result confirms that buffering is beneficial for small number of attributes.
When the number of attributes increases to a few thousand per instance, the performance of these more elaborate algorithms drops considerably.
However, the \vht \wok continues to perform relatively well and better than sharding.
This performance, coupled with good speedup over \moa (as shown next) makes it a viable option for streams with a large number of attributes and a large number of instances. 

\begin{figure}
\centering
\includegraphics[scale=0.8]{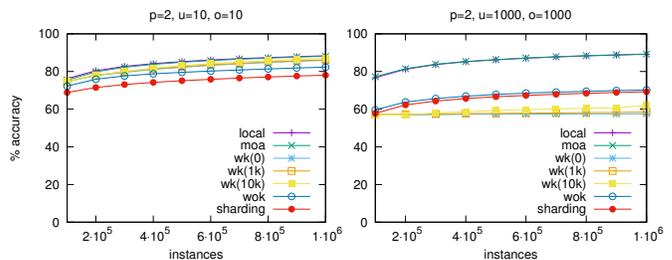}
\caption{Evolution of accuracy with respect to instances arriving, for several versions of \vht (local, \wok, \wkz) and sharding, for dense datasets.}
\label{fig:dense_accuracy_evolution}
\end{figure}

\begin{figure}
\centering
\includegraphics[scale=0.8]{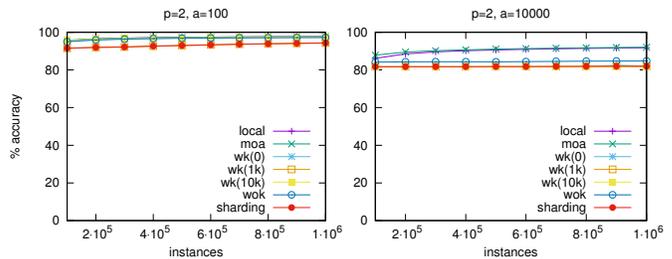}
\caption{Evolution of accuracy with respect to instances arriving, for several versions of \vht (local, \wok, \wkz) and sharding, for sparse datasets.}
\label{fig:sparse_accuracy_evolution}
\end{figure}

\subsection{Speedup of \vht distributed vs. \moa}

Since the accuracy of \vht \wkz is not satisfactory for both types of instances, next we focus our investigation on \vht \wok.

Figure~\ref{fig:dense_speedup_withmoa} shows the speedup of \vht for dense instances.
\vht \wok is about 2-10 times faster than \vht local and up to 4 times faster than \moa.
Clearly, the algorithm achieves a higher speedup when more attributes are present in each instance, as ($i$) there is more opportunity for parallelization, and ($ii$) the implicit load shedding caused by discarding instances during splits has a larger effect.
Even though sharding performs well in speedup with respect to \moa on small number of attributes, it fails to build a model for large number of attributes due to running out of memory.
In addition, even for a small number of attributes, \vht \wok outperforms sharding with a parallelism of $8$.
Thus, it is clear from the results that the vertical parallelism used by \vht offers better scaling behavior than the horizontal parallelism used by sharding.

\begin{figure*}
\centering
\includegraphics[scale=0.8]{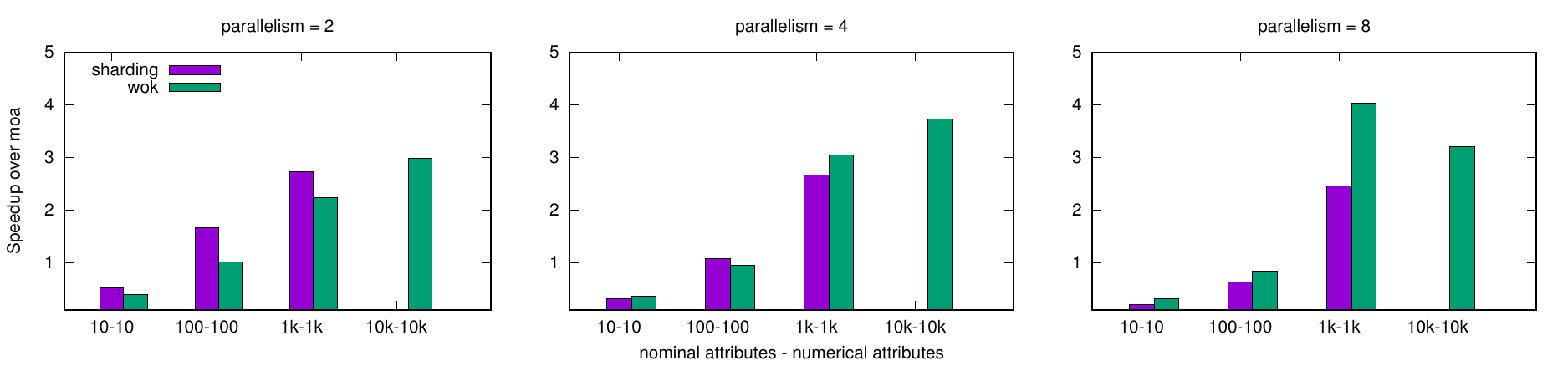}
\caption{Speedup of \vht \wok executed on \samoa compared to \moa for dense datasets.}
\label{fig:dense_speedup_withmoa}
\end{figure*}

When testing the algorithms on sparse instances, as shown in Figure~\ref{fig:sparse_speedup_withmoa}, we notice that \vht \wok can reach up to 60 times the throughput of \vht local and 20 times the one of \moa (for clarity we only show the results with respect to \moa).
Similarly to what observed for dense instances, a higher speedup is observed when a larger number of attributes are present for the model to process.
This very large superlinear speedup ($20 \times$ with $p=2$), is due to the aggressive load shedding implicit in the \wok version of \vht.
The algorithm actually performs consistently less work than the local version and \moa.

However, note that for sparse instances the algorithm processes a constant number of attributes, albeit from an increasingly larger space.
Therefore, in this setup, \wok has a constant overhead for processing each sparse instance, differently from the dense case.
\vht \wok outperforms sharding in most scenarios and especially for larger numbers of attributes and larger parallelism.

Increased parallelism does not impact accuracy of the model (see Figure~\ref{fig:dense_accuracy_withlocal} and Figure~\ref{fig:sparse_accuracy_withlocal}), but its throughput is improved.
Boosting the parallelism from 2 to 4 makes \vht \wok up to 2 times faster.
However, adding more processors does not improve speedup, and in some cases there is a slowdown due to additional communication overhead (for dense instances).
Particularly for sparse instances, parallelism does not impact accuracy which enables handling large sparse data streams while achieving high speedup over \moa.

\begin{figure*}
\centering
\includegraphics[scale=0.8]{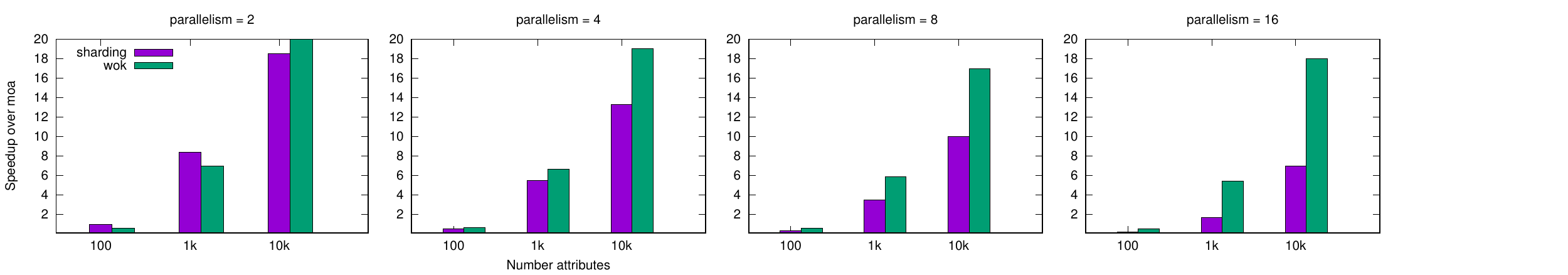}
\caption{Speedup of \vht \wok executed on \samoa compared to \moa for sparse datasets.}
\label{fig:sparse_speedup_withmoa}
\end{figure*}

\subsection{Performance on real-world datasets}

Tables~\ref{tab:real-accuracy} and~\ref{tab:real-time} show the performance of \vht, either running in a local mode or in a distributed fashion over a storm cluster of a few processors.
We also test two different versions of \vht: \wok and wk(0).
In the same tables we compare \vht's performance with \moa and sharding.

The results from these real datasets demonstrate that \vht can perform similarly to \moa with respect to accuracy and at the same time process the instances faster.
In fact, for the larger dataset, covtypeNorm, \vht \wok exhibits 1.8 speedup with respect to \moa, even though the number of attributes is not very large (54 numeric attributes).
\vht \wok also performs better than sharding, even though the latter is faster in some cases.
However, the speedup offered by sharding decreases when the parallelism level is increased from 2 to 4 shards.

\begin{table}[htbp]
\caption{Average accuracy (\%) for different algorithms, with parallelism level (p), on the real-world datasets.}
\vspace{-\baselineskip}
\begin{center}
\small
\tabcolsep=0.5em
\begin{tabular}{l r r r r r r r r}
\toprule
dataset	&	\moa	&	\multicolumn{5}{c}{\vht}	&	\multicolumn{2}{c}{Sharding}		\\
		&		&	local	&	\wok	&	\wok	&	wk(0)&	wk(0)&		&		\\
		&		&		&	p=2	&	p=4	&	p=2	&	p=4	&	p=2	&	p=4	\\
\midrule
elec		&	75.4	&	75.4	&	75.0	&	75.2	&	75.4	&	75.6	&	74.7	&	74.3	\\
phy		&	63.3	&	63.8	&	62.6	&	62.7	&	63.8	&	63.7	&	62.4	&	61.4	\\
covtype	&	67.9	&	68.4	&	68.0	&	68.8	&	67.5	&	68.0	&	67.9	&	60.0	\\
\bottomrule
\end{tabular}
\end{center}
\label{tab:real-accuracy}
\end{table}%

\begin{table}[htbp]
\caption{Average execution time (seconds) for different algorithms, with parallelism level (p), on the real-world datasets.}
\vspace{-\baselineskip}
\begin{center}
\small
\tabcolsep=0.5em
\begin{tabular}{l r r r r r r r r}
\toprule
Dataset	&	\moa		&	\multicolumn{5}{c}{\vht}	&	\multicolumn{2}{c}{Sharding}									\\
		&			&	local		&	\wok		&	\wok		&	wk(0)	&	wk(0)	&			&			\\
		&			&			&	p=2		&	p=4		&	p=2		&	p=4		&	p=2		&	p=4		\\
\midrule
elec		&	1.09		&	1		&	2		&	2		&	2		&	2		&	2		&	2.33		\\
phy		&	5.41		&	4		&	3.25		&	4		&	3		&	3.75		&	3		&	4		\\
covtype	&	21.77	&	16		&	12		&	12		&	13		&	12		&	9		&	11		\\
\bottomrule
\end{tabular}
\end{center}
\label{tab:real-time}
\end{table}%

\subsection{Summary}

In conclusion, our \vht algorithm has the following performance traits.
We learned that for a small number of attributes, it helps to buffer incoming instances that can be used in future decisions of split.
For larger number of attributes, the load in the model can be high and larger delays can be observed in the integration of the feedback from the local statistics into the model.
In this case, buffered instances may not be used on the most up-to-date model and this can penalize the overall accuracy of the model.

With respect to a centralized sequential tree model (\moa), it processes dense instances with thousands of attributes up to $4\times$ faster with only $10-20\%$ drop in accuracy.
It can also process sparse instances with thousands of attributes up to $20\times$ faster with only $5-10\%$ drop in accuracy.
Also, its ability to build the tree in a distributed fashion using tens of processors allows it to scale and accommodate thousands of attributes and parse millions of instances.
Competing methods cannot handle these data sizes due to increased memory and computational complexity.

\section{Conclusion}
\label{sec:conclusion}

The rapid increase observed in the number of users of social media or of IoT devices has lead to a multifold increase in the data available for analysis.
For these data to be analyzed and business or other learnings to be extracted, new machine learning methods are required, able to scale to a large size of fast data arriving at high speeds.

In this paper we presented the Vertical Hoeffding Tree (\vht), the first distributed streaming algorithm for learning decision trees that can be used for performing classification tasks on such large data streams arriving at high rates.
\vht features a novel way of distributing decision trees via vertical parallelism.
The algorithm is implemented on top of \asamoa, a platform for mining big data streams, and is thus able to run on real-world clusters.

Through exhaustive experimentation, and in comparison to a centralized sequential tree model, we show that \vht can process dense and sparse instances with thousands of attributes up to $4\times$ and $20\times$ faster, respectively, and with small degradation in accuracy.
Also, \vht's ability to build the decision tree in a distributed fashion using tens of processors allows it to scale and accommodate thousands of attributes and parse millions of instances.
We also show that competing distributed methods cannot handle the same data sizes due to memory and computational complexity.

\balance

\bibliographystyle{nourlabbrvnat}
\bibliography{references}
\end{document}